# Thermodynamic and ordering kinetics in asymmetric PS-*b*-PMMA block copolymer thin films

Gabriele Seguini,[1,*] Fabio Zanenga,[1] Gianluca Cannetti,[1] Michele Perego,[1]

[1] *IMM-CNR, Unit of Agrate Brianza, Via C. Olivetti 2, I-20864 Agrate Brianza, Italy.*

***ABSTRACT***

The ordering kinetics of standing cylinder-forming polystyrene-*block*-poly (methyl methacrylate) block copolymers (molecular weight: 39 kg/mol) close to the order-disorder transition is experimentally investigated following the temporal evolution of the correlation length at different annealing temperatures. The grow exponent of the grain coarsening process is determined to be 1/2, signature of a curvature driven ordering mechanism. The measured activation enthalpy and the resulting Meyer-Neldel temperature for this specific copolymer along with the data already known for PS-*b*-PMMA block copolymers in strong segregation limit, allows investigating the interplay between the ordering kinetics and the thermodynamic driving force during the grain coarsening. These findings unveil various phenomena concomitantly occurring during the thermally activated ordering kinetics at segmental, single chain, and collective level.

---

(GS) gabriele.seguini@mdm.imm.cnr.it

***Introduction.*** Block copolymers (BCPs) are a single component system composed by two covalently bonded and repelling macromolecules. The chain connectivity frustrates the macro phase separation arising from block repulsion. Low temperature, low symmetry, spatially modulated phases self-assembly upon phase separation of the homogenous phase. The equilibrium morphologies, such as lamellae, cylinders, gyroids, and spheres, are mainly determined by the relative volume fraction of one component quantified by the chain disproportion ($f$). The degree of polymerization, *i.e.* the number of monomeric units ($N$), dictates the domain spacing ($L_0$) characterizing the periodicity of the self-assembled domains. The change in local free energy due to the interaction between the two distinct monomers is quantified by the Flory-Huggins parameter ($\chi$). [1], [2], [3], [4], [5], [6], [7], [8], [9]

The BCP free energy is composed by nearly specular entropic ($\div 1/N$) and enthalpic ($\div \chi$) contributions. [4] At the order-disorder transition (ODT), these contributions are balanced. The segregation strength ($\chi N$) determines the strength of the repulsive interaction between the two blocks. $\chi N$ governs the phase separation as well as the degree of segregation between the two blocks upon phase separation. Increasing $\chi N$ above $\chi N_{\mathrm{ODT}}$, the transition to the ordered phase is driven by the free energy gain associated to the phase separation. This free energy gain results from the balance between the enthalpic gain and the entropic penalty in the transition to the phase-separated state from the homogeneous phase. The potential describing the phase separation in BCP self-assembly is short-range attractive and long-range repulsive (SALR) along the backbone of the polymer chain. [3] In particular, the long-range repulsion has entropic origin. [1] The entropic penalty contributions, such as the chain stretching and the interface localization between the incompatible polymeric chains, determine the polymeric chain features of the phase-separated state. The chain stretching increases while the interface localization narrows increasing $\chi N$ and moving away from $\chi N_{\mathrm{ODT}}$. [4], [9], [10], [11]

Above ODT, the degree of segregation between the two blocks is usually categorized in three regimes: weak, intermediate, and strong according to the concentration gradient across the interface between microphase separated domains as a function of $\chi$ and N. The weak segregation regime occurs close to ODT ($\chi N \approx \chi N_{ODT}$) when the composition profile is almost sinusoidal. Conversely, the strong segregation regime takes place far from the ODT ($\chi N > \chi N_{ODT}$) and is characterized by a nearly flat composition profile as well as strongly stretched chains and narrow interfaces. The weak to intermediate crossover is commonly placed slightly above ODT before the onset of the strong segregation. However, considering a mean field based approach, strong segregation conditions have been demonstrated valid over most of the intermediate regime. Overall, the segregation regimes can be classified as weak (WSL) and strong (SSL) segregation limit with a smooth transition between the two limits. The degree of segregation in the two regimes can be effectively characterized by $L_0$ and the interfacial width, $w$: in particular $w/L_0 \approx 1$ in the WSL while $w/L_0 << 1$ with $L_0 \approx N^{2/3}$ in the SSL. [9], [10], [11], [12]

Phase diagram properties for BCP system at thermodynamic equilibrium have been widely investigated. [3], [10], [13] However, the final equilibrium state is not reached instantaneously and the kinetics of equilibrium structure formation proceed through the coarsening of the polycrystalline pattern by progressive annihilation of the topological defects and reorientation of ordered domains. Up to now, the major focus has been on the dynamics of the coarsening process. [13], [14], [15], [16] Ordering kinetics has been described in the framework of nucleation and growth of ordered mesophases [17], [18], [19] as well as focusing on the reduced chain diffusivity increasing the chain length. [20], [21] However, an in depth experimental investigation of the relationship between the thermodynamic driving force and the ordering kinetics is still missing.

The weakly dependence on the temperature ($T$) of the Flory-Huggins parameter $\chi_{S\text{-}MMA}$ for a polystyrene-*block*-poly (methyl methacrylate) (PS-*b*-PMMA) BCP defines a peculiar experimental configuration: the speed of the ordering kinetics is modulated by changing the annealing temperature $T$ without affecting the thermodynamic driving force $\chi N$ that is independently defined by $N$ of the selected copolymer. This peculiarity allows investigating the specific interplay between the ordering kinetics and the thermodynamic driving force during the coarsening process. Recently, within this experimental approach, cylinder-forming PS-*b*-PMMA BCP thin films in SSL have been scrutinized. [14] Further reducing $N$ of the investigated BCP allows approaching ODT, where the system is thermodynamically less stable, and reducing $L_0$ toward the lower limit. [22] Moreover, reduced $L_0$ and effective ordering are crucial for the exploitation of BCPs as valuable technological platform for fabrication of dense and ordered nanostructures. All these fundamental and practical characteristics make such a system very interesting to be scrutinized.

In this work, we investigate the thermally activated ordering for a cylinder forming PS-*b*-PMMA BCP with $\chi N \approx \chi N_{ODT}$ measuring the evolution of the level of order as a function of the different time and temperature combinations of the thermal treatment. The data herein reported allow a comprehensive clarification of the kinetic and thermodynamic mechanisms occurring at collective, single chain, and segmental levels, in the ordering process of a BCP based two-dimensional hexagonal pattern for all the segregation regimes. [16], [23], [24], [25], [26]

***Experimental.*** Poly (styrene-*random*-methyl methacrylate) (P(S-*r*-MMA), molecular weight, $M_n$=1.7 and 69 kg/mol, $f_s$=0.61, polydispersity index, PDI=1.19) random copolymer (RCP) was grafted by means of a rapid thermal annealing (RTA) ($T$=310 °C, $t$=60 s) on hydroxyl terminated, flat, and unpatterned Si surface to form a neutral brush layer. This surface neutralization induces the perpendicular orientation of the PMMA cylinders in the self-assembled asymmetric BCPs. PS-*b*-PMMA BCPs with styrene fraction, $f_S$=0.69, PDI=1.07, and $N$=379 ($M_n$=39 kg/mol) in toluene solution were spinned on the RCP grafted layer to obtain a ≈30 nm BCP thick layer. The spin coated films underwent a thermal treatment in RTA in order to enhance the polymeric chain mobility and to drive the system toward the ordered equilibrium morphology. [22], [27], [28] The ordering process in the BCP thin films was studied considering different combinations of annealing time ($t_A$=10-900 s) and temperature ($T_A$=160-220 °C). Surface morphology after RTA treatment was investigated by scanning electron microscopy (SEM). After SEM images binarization, the centroid of every cylinder was localized by software analysis and a Delaunay triangulation was performed to detect the coordinates of the vertex of the hexagons. These analyses allowed investigating the pattern orientation in real space. The level of order of the two-dimensional hexagonal pattern was quantified by measuring the correlation length ($\xi$) fitting the resulting autocorrelation function with an exponential decay function. This method provides information over an area that is much larger than the measured $\xi$. [22], [27], [28], [29], [30]

***Results and Discussion.*** During the deposition process by spin casting from a dilute solution the BCP thin films reach a kinetically trapped state far from the thermodynamic equilibrium state due to the slow dynamic of the polymeric chains. [27], [30] A thermal or solvent annealing process is necessary to drive the system from this metastable state towards the thermodynamic equilibrium. Recently we demonstrated that the grafted layer of RCP, that is commonly used to neutralize the surface, can trap a wealth of residual solvent in the polymeric film. This reservoir of solvent can be exploited during the thermal annealing to promote phase separation and grain coarsening in thin films of PS-*b*-PMMA in the SSL regime. [8], [14] The capability of the RTA treatment to reach the target temperature in very short time, even though not instantaneous, reduces solvent evaporation during the heating step and maximizes the effect of the residual solvent trapped in the polymeric film to enhance the polymeric chain mobility. [14], [27], [28] It has been evidenced that the amount of residual solvent in the overall RCP+BCP system is proportional to the brush layer thickness ($h$) that, indeed, is proportional to the RCP $M_n$. [27], [30]

The effect of the residual solvent on the self-assembly process of a thin film of PS-*b*-PMMA with low $N$ (=379) have been investigated. Figure 1 shows the SEM plan view images of the self-assembled BCP over a "*thick*" brush layer ($M_n$(RCP)=69 kg/mol, $h$≈19 nm). [27] The investigated surfaces are phase separated but poorly ordered irrespective of the specific combination of $T_A$ and $t_A$. In these samples, even in the best conditions, $\xi$ values are very low and it is not possible to detect a valuable ordering kinetics as a function of ($t_A$, $T_A$). This fact evidences that for a BCP system with low $N$, an excessive amount of solvent in the overall RCP+BCP system dilutes the investigated BCP system, shielding the block interaction forces, reducing the segregation strength, and driving the BCP towards the homogenous phase. Differently, using a "*thin*" brush layer ($M_n$(RCP)=1.7 kg/mol, $h$≈2 nm) as neutralizing layer the reduced amount of solvent makes possible to achieve the phase separation along with a valuable level of order as evidenced by SEM images reported Fig. 2. [14], [27] In particular, Figure 2 shows that, for very short processing times ($t_A$=1-10 s) and relatively low temperatures ($T_A$=160-180 °C), lateral order is very limited (gray SEM images with $\xi<L_0$). Increasing $T_A$ (≥180 °C) and/or $t_A$ (≥10 s), the level of order increases with $\xi$ values well above $L_0$ ($1<\xi<2L_0$ blue, $\xi>2L_0$ green SEM images). Finally, for $T_A$=220 °C and $t_A$=900 s, the BCP surface shows inhomogeneities, suggesting the occurrence of a degradation process in the polymer film (red SEM images). The periodicity of the self-assembled domains was evaluated to be $L_0$=24.0±1.0 nm and the diameter of the hexagonal packed cylinders was determined to be $d$=12.0±2.0 nm. [22], [27] These results are in agreement with data previously collected on the same BCP and demonstrate that the low $N$ of the PS-*b*-PMMA corresponds to a $\chi N$ value that is very close to ODT, providing a signature of WSL. [22], [27]

All the following data were obtained by self-assembling the BCP film over a "*thin*" RCP layer with a RTA treatment in a range of temperatures between 180 and 200 °C. Figure 3 reports $\xi$ as a function of $t_A$ at different $T_A$ (180, 190, 195, 200 °C). Data evidence a regular increase of $\xi$ as a function of $t_A$ at the different $T_A$. While for the same $t_A$, $\xi$ increases increasing $T_A$. This regularity induces to treat the measured $\xi$ values applying the time-temperature superposition (TTS) procedure. $\xi$ as a function of $t_A$ for different $T_A$ are collected in a master curve reporting $\xi$ versus an equivalent time ($t_{EQ}$) at the reference temperature $T_A$=180 °C (Fig. 4). The master curve results by multiplying for a proper rigid shift factor the time scale of $\xi(t_A)$ curves for every $T_A$>180 °C in order to match $\xi$ values to the reference curve at $T_A$=180 °C. [14] The master curve is well fitted with a power law $t^{\phi}$, evidencing the self-similarity of the system. The measured growth exponent is $\phi$=0.50±0.02.

The evolution of $ln\xi$ as a function of $1/T_A$ at different $t_A$ is reported in Fig. 5. Experimental data were fitted using an Arrhenius equation $ln\xi=lnA-H_A/(K_B\cdot T_A)$, where $H_A$ corresponds to the activation enthalpy of the grain coarsening process, $K_B$ is the Boltzmann constant, and $A$ is the pre-exponential factor, that is related to the activation entropy $S_A$. The linear evolution of $ln\xi$ as a function of $1/T_A$ indicates that the grain coarsening process is a thermally activated process with a kinetic barrier. Limited variations of $H_A$ with $t_A$ have been observed, as previously reported also in the case of PS-*b*-PMMA in SSL. [14] The mean $H_A$ value is 136.5±5.0 kJ/mol.

These data for a PS-*b*-PMMA BCP close to ODT along with the previously published data for PS-*b*-PMMA BCP in SSL allow a comprehensive description of the system. [14] In particular, considering the BCP ordering process as an outcome of various phenomena concomitantly occurring at different levels (collective, single chain, segmental) it is possible to map out a profitable route for the rationalization of the experimental data.

*Collective dynamics.* In Fig. 6 (data in SSL from ref. [14]) $\phi$ values for BCP with 39<$M_n$<132 kg/mol are depicted as a function of $N_{MMA}$, *i.e.* the number of monomeric units in the minor component of the BCP. The absolute $\phi$ value progressively increases when decreasing $N_{MMA}$. In particular, for the BCPs in SSL with $f_S$=0.69, $\phi$ exhibits an analytical dependence $exp(-\chi N_{MMA})$, with $\chi$=0.028±0.002. This effective $\chi$ parameter is in excellent agreement with $\chi_{S\text{-}MMA}$ values reported in the literature. [14], [31] Moreover, the exponential dependence of $\phi$ on $\chi N_{MMA}$ mimics the one of a single chain diffusion, suggesting that this mechanism drives the grain coarsening process. For the 39 kg/mol BCP ($N_{MMA}$=122, $f$=0.69), *i.e.*

when entering in WSL, $\phi$ value further increases but with a value which is lower than compared to what expected according to the extrapolated exponential curve. This dissimilarity evidences that, approaching ODT, a supplemental rate limiting mechanism frustrates the diffusion limited grain coarsening mechanism observed in SSL.

This frustration can be rationalized as follows. Approaching ODT, the chains are less stretched and the interface between PS and PMMA components is not sharp, hindering the expected accelerated kinetics associated to the increased diffusivity of BCP chains when decreasing $N$. [15], [32] Growth exponents of 1/2 and 1/3 are widely recognized as maximum values for curvature driven and diffusion limited coarsening, respectively, for systems with short-range attractive potential that macro phase separate. [33], [34], [35], [36] In the data herein reported, the same maximum growth exponents are experimentally measured for a system described by a SALR potential inducing the microphase separation and the inherent thermally activated ordering. It is worth noticing the subtle dependence of these data on $f_S$. In fact, the $\phi$ value that was obtained for the 48 kg/mol BCP ($N_{MMA}$=125, $f_S$=0.74) is clearly outside this exponential trend. This is ascribed to the dependence of the self-diffusion coefficient on 1/$N$, further supporting the interpretation of these data as a marker of a diffusion limited mechanism.

*Single chain kinetics and thermodynamic driving force coupling.* Figure 7 compares the $H_A$ value herein achieved with those previously reported in the case of BCP in SSL. [14] $H_A$ values are reported as a function of $N_{MMA}$, and follow the same exponential decay $exp(-\chi N_{MMA})$ irrespective of the segregation regime. The $\chi$ value obtained from the overall fitting is 0.024±0.001, in excellent agreement with the previously reported value in the case of SSL ($N_{MMA} \geq 167$, $\chi$=0.024) and with $\chi_{S\text{-}MMA}$ values reported in the literature. [14], [31]

This analysis highlights that the energetic associated with the thermally activated grain coarsening process is the same in WSL and SSL. Indeed, the kinetic barrier evolution is described by the same exponential dependence on -$\chi N_{MMA}$ in both regimes. This feature suggests the same basic process irrespective of $N$. It can be depicted as a hindered diffusion mechanism similar either to the single chain hopping in sphere based systems or to the perpendicular diffusion in lamellar systems. [32], [37], [38], [39], [40] Actually, the two-dimensional hexagonally packed lattice of the BCP film is represented as a system with zero dimensional connectivity where the energy penalty for hopping diffusion describes the experimental data.

The effect of the thermodynamic driving force on the kinetic process can be rationalized considering the specular mechanisms of chain unmixing and mixing. The free energy gain in unmixing corresponds to a free energy penalty in mixing. Experimentally, phase separation and ordering are two consecutive steps of the self-assembly process; early stage phase separation is assumed to take place when $\xi \leq L_0$ and late stage ordering when $\xi > L_0$. [14], [27] However, the same thermodynamic driving force $\chi N$ determines nucleation and growth during phase separation at the expense of the homogenous phase ($\xi \leq L_0$) and coarsening of the ordered phase at the expense of a less ordered phase during ordering ($\xi > 2L_0$). [17], [18], [19] In this respect, the framework of nucleation accounts for the herein reported experimental data. Nucleation corresponds to a thermally activated transition from a less stable to a more stable state, moving through a transition state at high energy, which represents the nucleation barrier. [41], [42], [43], [44], [45] Indeed, the basic features of nucleation are a free energy variation, with competing gain and penalty components, along with a nucleation barrier that inversely depends on the free energy gain during nucleation. All these features match our experimental results. Indeed, in their seminal work about the kinetics of metastable states in BCPs, Fredrickson and Binder disclosed that "*the barrier for homogeneous nucleation in symmetric diblocks decreases with increasing molecular weight*" [17] in agreement with our experimental findings. Alternatively, a theoretical approach based on single-chain in mean field simulations evidenced peculiarities that describe our experimental results as well. In particular, focusing on the local densities of the blocks as collective variables, such a theoretical approach evidences that the slowness of the ordering kinetics originates not from a collective variable but from a small Onsager coefficient, that connects the thermodynamic driving force to the order parameter flux. Within this approach, the exponential dependence of the barrier on $-\chi N_{MMA}$ is reported to be related to the single chain dynamic without a collective free energy barrier in the minimum free energy path. [37], [39], [46] Indeed, Müller and de Pablo reported that "*the kinetics can be protracted in strongly segregated systems, even if there is a strong thermodynamic driving force*", consistently with our observations showing that increasing the thermodynamic driving force the kinetics gets slower. [46]

*Segmental activation.* Figure 8 reports the relationship between $H_A$ and $lnA$, *i.e.* $ln$ of the pre-exponential factor $A$ of the Arrhenius equation that is related to the activation entropy $S_A$. Present data are compared with those obtained in SSL from ref. [14]. The overall linear relationship indicates that data follow the Meyer-Neldel (MN) rule $\Delta S=\Delta H/T_{MN}$, where $T_{MN}$ is the MN temperature. [47], [48] From the linear fitting, $T_{MN}$ results to be 447 K (174 °C). Considering the reported mean bulky glass transition temperature $T_G$ for these BCPs, it descends that $T_{MN} \approx 1.2\ T_G$. This fact evidences the same behavior in SSL and WSL, indeed $T_{MN} \approx 1.2\ T_G$ in both cases.

The same $T_{MN}$ value, irrespective of WSL or SSL, refers the ordering kinetics to a thermally activated process that begins at the same temperature without any dependence on $N$. The value of $T_{MN} \approx 1.2\ T_G$ ascribes this thermally activated process to the glass transition (GT) where segmental dynamics, which is the basic kinetic step, becomes active. Above this threshold, the inverse relationship between diffusivity and viscosity holds. This segmental feature originates the collective TTS behavior of BCP systems that has been exploited following the ordering kinetics. [49], [50] From the thermodynamic point of view, it has been shown that $T_{MN}$ is a consequence of a balance between repulsive and attractive interactions. [51] This balance at the kinetic GT mimics the balance between entropic and enthalpic contributions at the thermodynamic ODT. [5]

*Overall view.* The experimental decouple of the kinetic and thermodynamic control evidences the two different phase transition (ODT and GT) that define BCP ordering. The distance from *ODT* ($N_{ODT}$) is determined by $N$. Considering that $\chi_{S\text{-}MMA}$ is almost independent on $T$, $N$ also dictates $\chi N$ that determines the thermodynamic driving force affecting phase separation and ordering. $T$ measures the distance from GT ($T_{MN}$) and it originates thermal fluctuations to activate the cross of the kinetic barriers. Overall, the ordering process is enclosed in these landmarks and its effective rate originates from the coupling of these thermodynamic and kinetics factors. As evidenced by the unique analytical dependence of $H_A$ and the single $T_{MN}$, thermodynamically, WSL and SSL are not distinguishable. On the other hand, approaching ODT, in WSL, the collective dynamic (quantified by $\phi$) is hindered by a supplemental rate limiting process that switch the ordering process from a diffusion limited (in SSL) to a curvature driven mechanism (in WSL).

The grain coarsening occurs by local rearrangement of single chains. Their mobility determines the cooperative diffusion (as quasi particles) and annihilation of topological defects that dominate the grain coarsening. [8], [32], [38] The apparent discrepancy of the trends of experimental kinetics barriers for the grain coarsening process (single chain based), decreasing with $N$, and the simulated free energy barriers for an overall defect removal, increasing with $N$, is recomposed considering that a collection of many molecules contributes to a defect. Both single chain and collective motions comply with the extensively evidenced kinetic limited coarsening and the reduced diffusivity, increasing $N$ and moving away from ODT, and, conversely, with the thermodynamic limited coarsening and the enhanced diffusivity, decreasing $N$ and moving towards ODT. [12], [21], [32], [37], [40], [46], [52]

Generally, an inverse coupling between kinetic barriers and thermodynamic driving force either in a two state kinetic model or in flux-force relationship is a ubiquitous concept in different scientific and technological fields such as metallurgy, protein folding, structural transitions, and metabolism. In particular, in the latter case, enzymes instead of stochastic forces (as thermal fluctuations) are used to activate the crossing of the barrier. [53], [54], [55], [56], [57], [58], [59], [60]

***Conclusions.*** From the measure of the evolution with time and temperature of a collective variable, the correlation length $\xi$, it was possible to experimentally evidence the basic thermodynamic that affects the ordering as well as the temperature where the polymer is not yet glassy identifying the onset of segmental activation. The coupling between kinetics and thermodynamic driving force has been comprehensively scrutinized by studying the ordering of a two-dimensional hexagonal pattern on a flat and unpatterned surface within a proper experimental PS-*b*-PMMA BCP based platform. These results are relevant both considering the paradigmatic properties of self-assembling BCP systems described by SALR competing interactions and the technological applications in order to plan better strategies to increase the order of dense nanostructures. [61], [62], [63], [64], [65], [66], [67], [68]

***Acknowledgements***

The authors acknowledge: M. Laus, K. Sparnacci, V. Gianotti, and D. Antonioli (Università Del Piemonte Orientale, Italy) for RCP synthesis; M. Alia (CNR-IMM, Italy) for assistance in the experimental part; and T. J. Giammaria (CNR-IMM, Italy, present address: CEA-LETI, France) and F. Ferrarese Lupi (CNR-IMM, Italy, present address: INRIM, Italy) for fruitful discussion. This research has been partially supported by the project IONS4SET" funded from the European Union's Horizon 2020 research and innovation program under Grant No. 688072.

## *FIGURES*

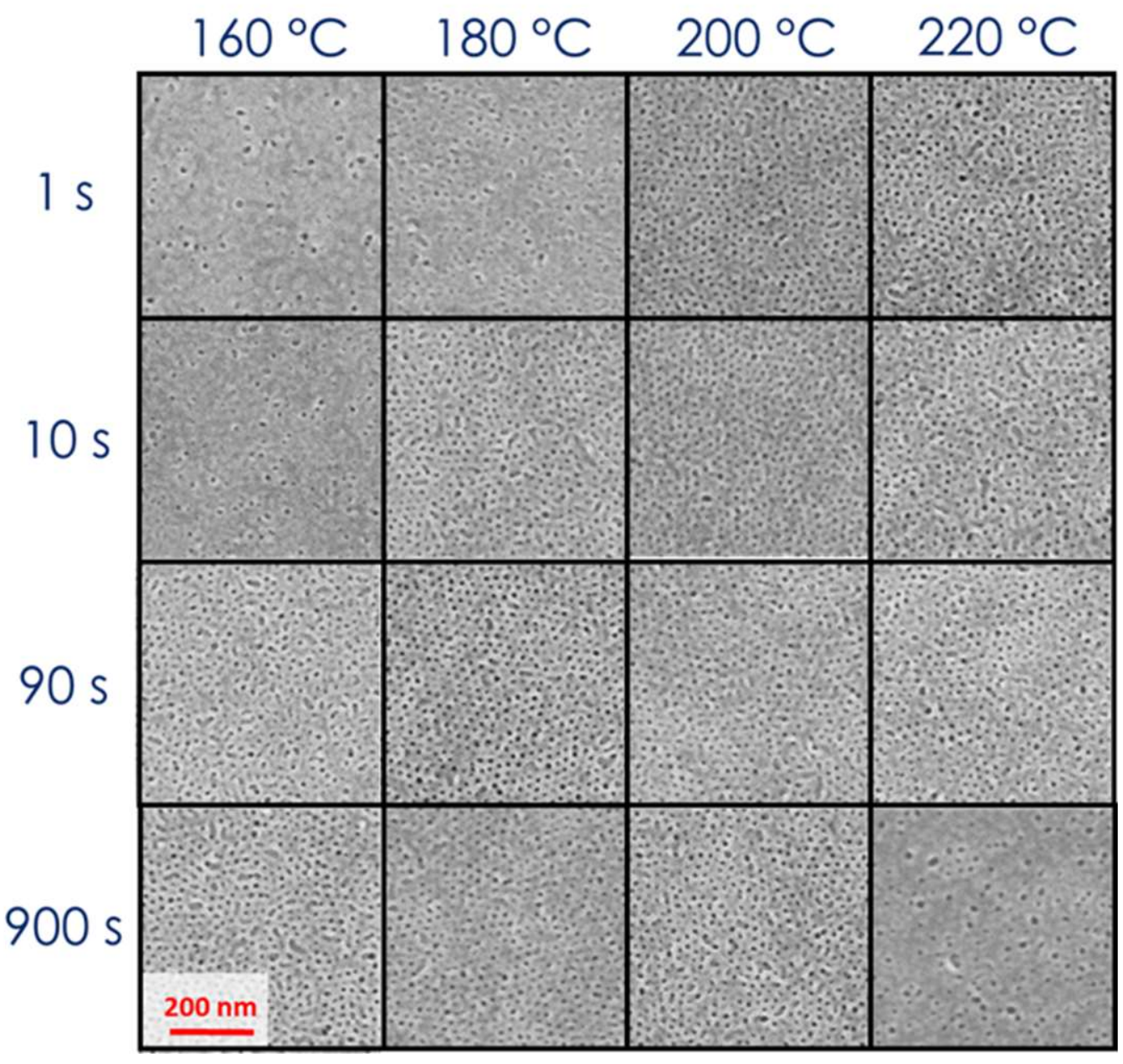

***Figure 1*** *Plan view high-magnification SEM images of 39 kg/mol BCP thin films annealed at different temperatures $T_A$ for different $t_A$ on "thick" RCP (69 kg/mol, brush layer ≈19 nm). $\xi<L_0$ gray images.*

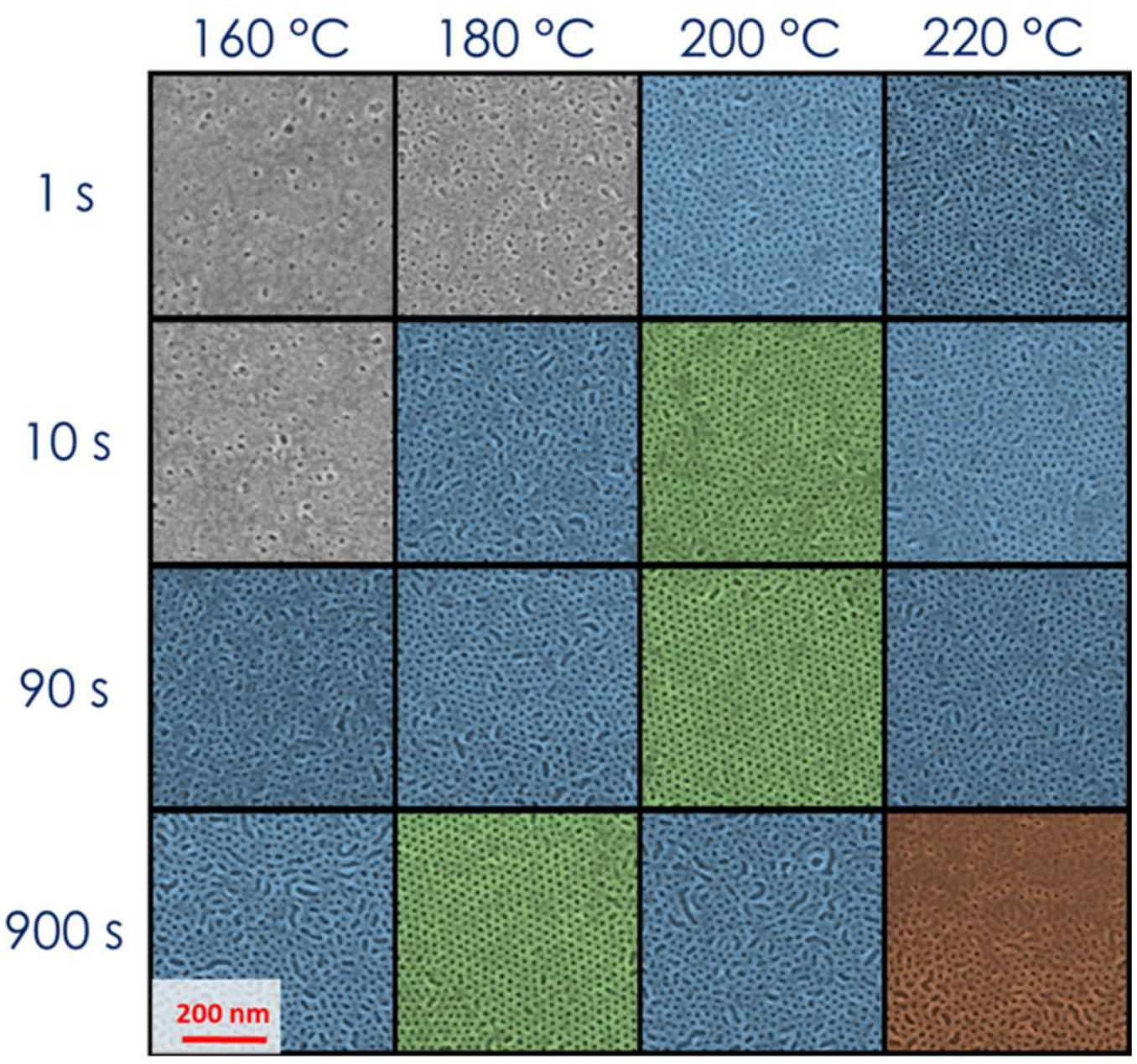

***Figure 2*** *Plan view high-magnification SEM images of 39 kg/mol BCP thin films annealed at different temperatures $T_A$ for different $t_A$ on "thin" RCP (1.7 kg/mol, brush layer ≈2 nm). $\xi<L_0$ gray images, $1<\xi<2L_0$ blue images, $\xi>2L_0$ green images, inhomogeneities red images.*

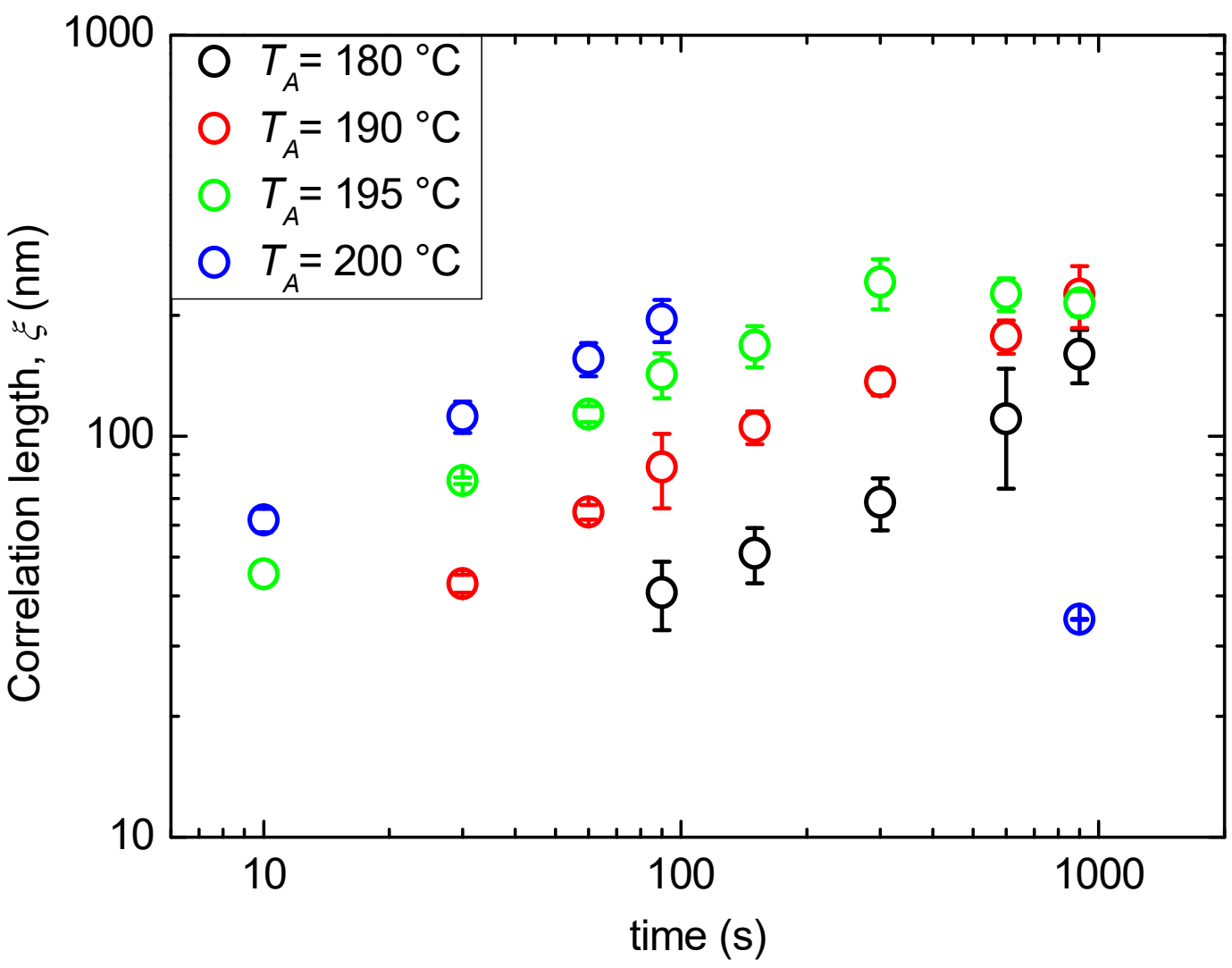

***Figure 3*** *Evolution of the correlation length ξ as a function of $t_A$ at different $T_A$ (180, 190, 195, 200 °C).*

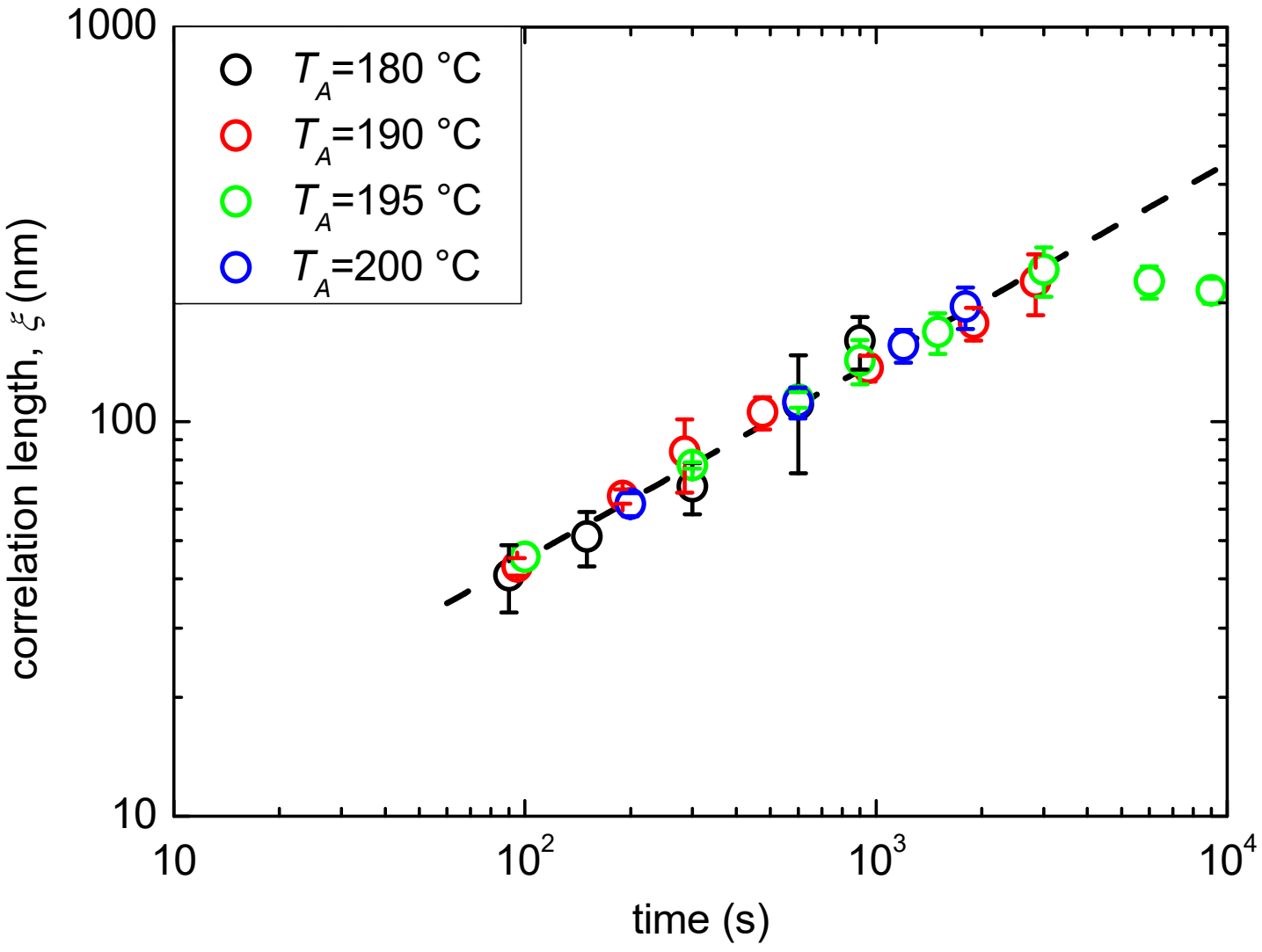

***Figure 4*** *Master curve from time–temperature-superposition of the correlation length data. The slope of the curve represents the growth exponent φ.*

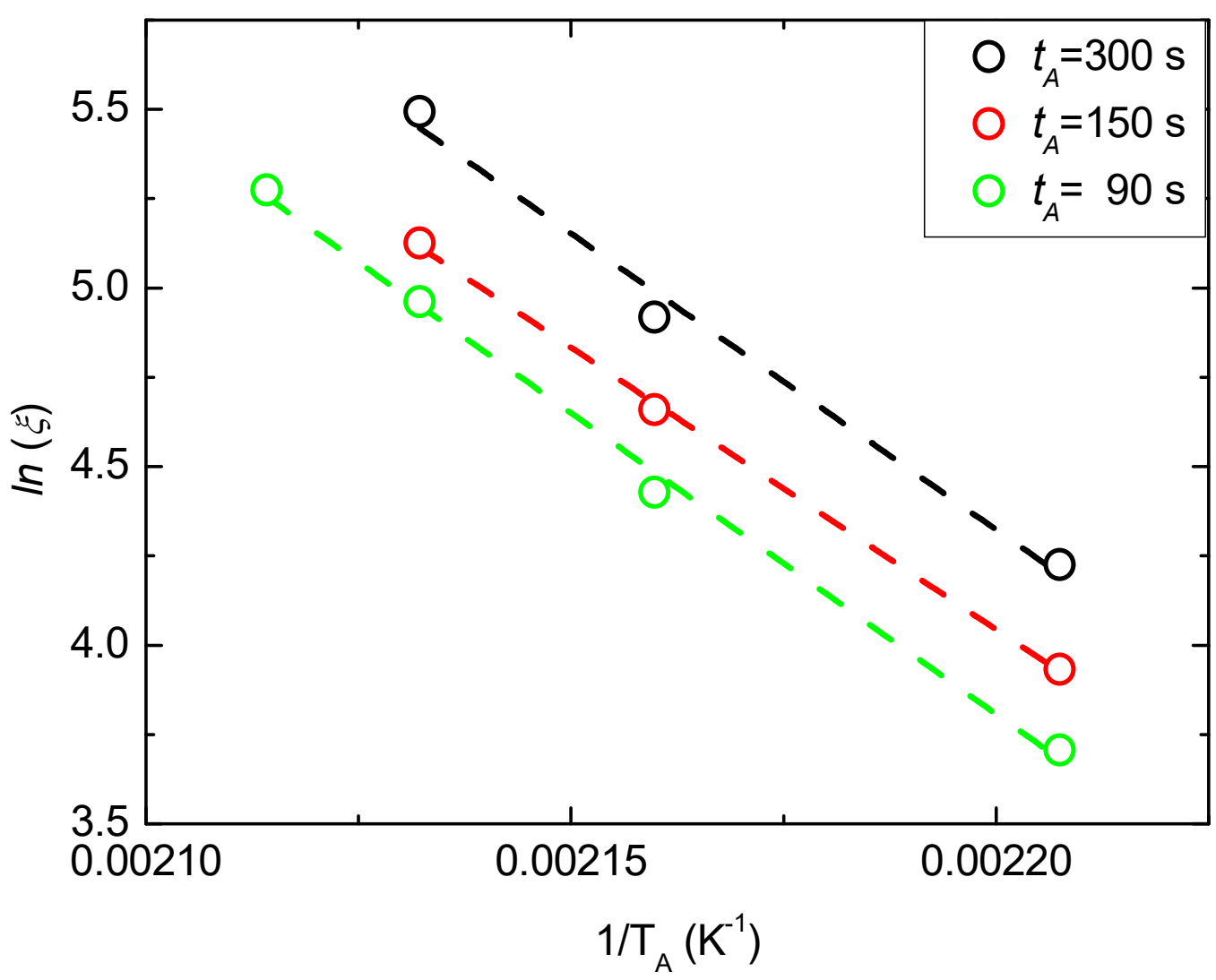

***Figure 5*** *Arrhenius plot of ξ data at different $t_A$.*

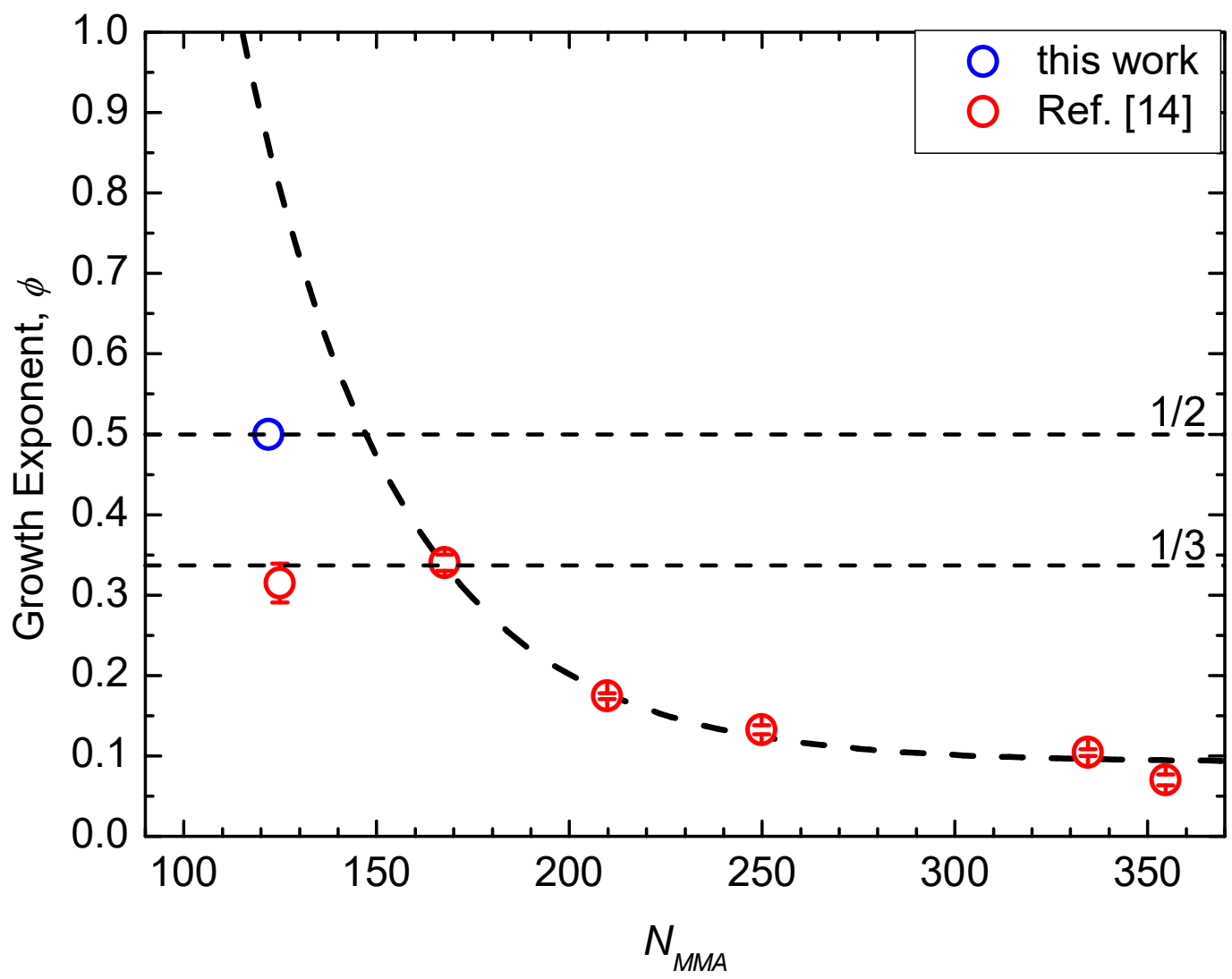

***Figure 6*** *Growth exponent φ as a function of $N_{MMA}$. Data in strong segregation limit ($N_{MMA}$≥167) are reported from ref.* [14].

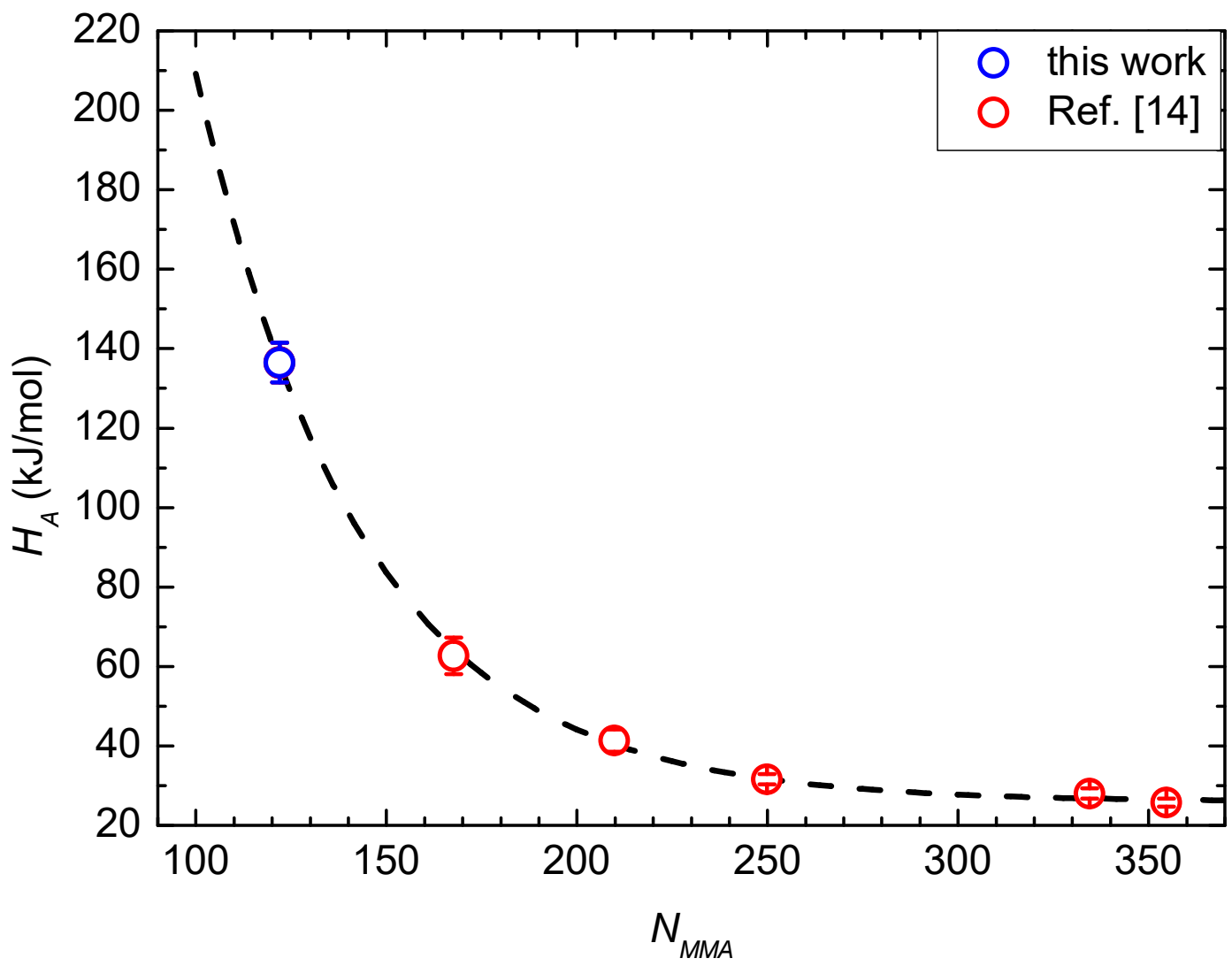

***Figure 7*** *Activation enthalpy $H_A$ as a function of $N_{MMA}$. Data in strong segregation limit ($N_{MMA}$≥167) are reported from ref.* [14].

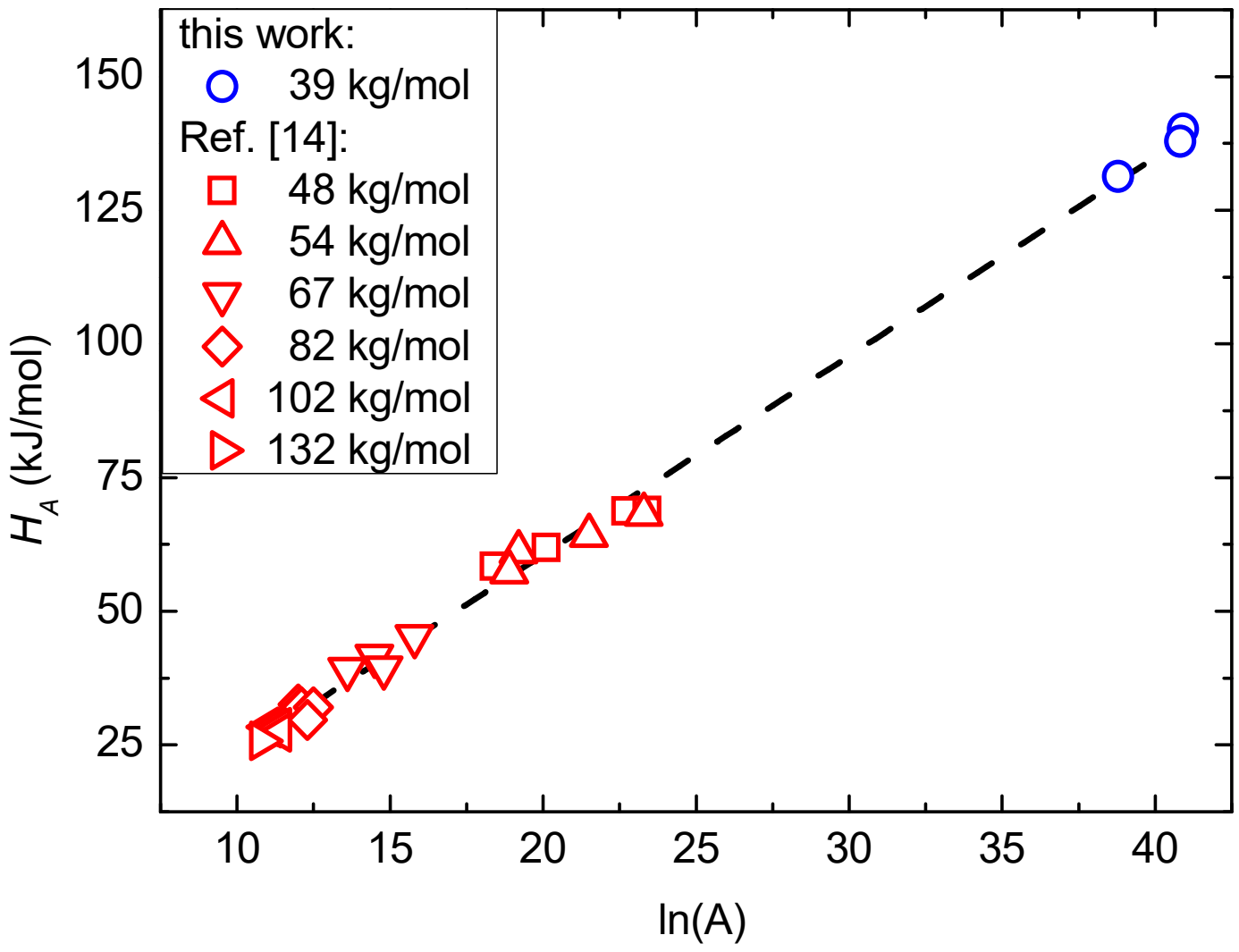

***Figure 8*** *$H_A$ as a function of the pre-exponential factor (lnA) of the Arrhenius plot corresponding to the activation entropy ($S_A$). Data in strong segregation limit are reported from ref.* [14].